\documentclass[review,3p,12pt]{elsarticle}
\usepackage[utf8]{inputenc}
\usepackage[T1]{fontenc}
\usepackage{graphicx}
\usepackage[export]{adjustbox}
\graphicspath{ {./images/} }
\usepackage{amsmath}
\usepackage{amsfonts}
\usepackage{amssymb}
\usepackage{mhchem}
\usepackage{stmaryrd}
\usepackage{hyperref}
\hypersetup{colorlinks=true, linkcolor=blue, filecolor=magenta, urlcolor=cyan,}
\journal{Journal of Fluids and Structures}

\begin{document}

\begin{frontmatter}

\title{Predicting the vascular adhesion of deformable drug carriers in narrow capillaries traversed by blood cells }

\author[iit]{A. Coclite}
\ead{alessandro.coclite@iit.it}

\author[dmmm]{G. Pascazio}
\ead{giuseppe.pascazio@poliba.it}

\author[dmmm]{M. D. de Tullio}
\ead{marcodonato.detullio@poliba.it}

\author[iit]{P. Decuzzi\corref{cor}}
\ead{paolo.decuzzi@iit.it}

\cortext[cor]{Corresponding author}

\address[iit]{Laboratory of Nanotechnology for Precision Medicine, nPMed, Fondazione Istituto Italiano di Tecnologia, Via Morego 30-16163,  Genova, Italy}

\address[dmmm]{Dipartimento di Meccanica, Matematica e Management, Politecnico di Bari, Via Re David 200 -- 70125 Bari, Italy}

\begin{abstract}
In vascular targeted therapies, blood-borne carriers should realize sustained drug release from the luminal side towards the diseased tissue. In this context, such carriers are required to firmly adhere to the vessel walls for a sufficient period of time while resisting force perturbations induced by the blood flow and circulating cells. Here, a hybrid computational model, combining a Lattice Boltzmann (LBM) and Immersed Boundary Methods (IBM), is proposed for predicting the strength of adhesion of particles in narrow capillaries (7.5 $\mu \mathrm{m})$ traversed by blood cells. While flowing down the capillary, globular and biconcave deformable cells ( $7 \mu \mathrm{m}$ ) encounter $2 \mu \mathrm{m}$ discoidal particles, adhering to the vessel walls. Particles present aspect ratios ranging from $0.25$ to $1.0$ and a mechanical stiffness varying from rigid $(\mathrm{Ca}=0)$ to soft $\left(\mathrm{Ca}=10^{-3}\right)$. Cell-particle interactions are quantitatively predicted over time via three independent parameters: the cell membrane stretching $\delta p$; the cell-to-particle distance $r$, and the number of engaged ligand-receptor bonds $N_{\mathrm{L}}$. Under physiological flow conditions $\left(\mathrm{Re}=10^{-2}\right)$, rigid particles are detached and displaced away from the wall by blood cells. This is associated with a significant cell membrane stretching (up to 10\%) and rapid breaking of molecular bonds $\left(t u_{\max } / \mathrm{H}<1\right)$. Differently, soft particles deform their shape as cells pass by, thus reducing force perturbations and extending the life of molecular bonds. Yet, only the thinnest deformable particles $(2 \times 0.5 \mu \mathrm{m})$ firmly adhere to the walls under all tested configurations. These results suggest that low aspect ratio deformable particles can establish long-lived adhesive interactions with capillary walls, enabling de facto vascular targeted therapies.
\end{abstract}

\end{frontmatter}

\section{Introduction}
Targeting the diseased vasculature is an attractive strategy for the diagnosis and treatment of a variety of pathologies. In cancer, endothelial cells express unique receptor molecules, such as integrins $\alpha_{\mathrm{v}} \beta_{3}$ and $\alpha_{\mathrm{v}} \beta_{5}$, tumor endothelial markers, vascular epidermal growth factor receptors, and so on (Neri and Bicknell (2005) and Atukorale et al. (2017)). In cardiovascular and chronic inflammatory diseases, the endothelium presents inflammatory molecules, such as P- and E-selectins, ICAM and VCAM-1, higher densities as compared to the normal vasculature (Libby, 2002; Ta et al., 2018; Soriano et al., 2000). Even, within the white adipose tissue, specific receptors are exposed on the surface of endothelial cells (Kolonin et al., 2004; Daquinag et al., 2011; Anselmo and Mitragotri, 2017). Although some of these receptors could be directly used as targets for small therapeutic molecules, most of them would serve as docking sites for blood-borne, vascular targeted particles (Kolhar. et al., 2013; Decuzzi and Ferrari, 2008). Engineered micro- and nano-carriers for the precise delivery of multiple agents are entering clinical trials for the diagnosis and treatment of a variety of diseases, including cancer, cardiovascular and neurological (Peer et al., 2007; Mulder et al., 2014). These carriers are realized using different techniques where attributes such as the size, the composition, the surface properties and, more recently, the shape and mechanical stiffness can be finely and independently tuned (Euliss et al., 2006; Godin et al., 2012; Muro et al., 2008; Palange et al., 2017; Palomba et al., 2018; Anselmo et al., 2015). Vascular targeted carriers can deposit at the luminal side large amounts of imaging and therapeutic molecules whose controlled release towards the diseased tissue can be triggered via a number of mechanisms, including chemical and mechanical stimuli (Mura et al., 2013).

Several reports have investigated the vascular transport of micro and nano-carriers in the attempt to identify the optimal configuration that could favor their deposition on endothelial walls. For instance, the works of Liu and collaborators showed that red blood cells would favor the lateral drifting and vascular binding of sufficiently large nanoparticles (Tan et al., 2011). Similarly, the authors demonstrated that the contribution of red blood cells is mostly amplified for micron-sized particles over conventional $100 \mathrm{~nm}$ nanoparticles (Lee et al., 2013). A more comprehensive analysis was provided by Vahidkhah and Bagchi, who considered the vascular transport and adhesion of spherical, prolate and oblate spheroids with differ aspect ratios (Vahidkhah and Bagchi, 2015). Their numerical results confirmed that oblate spheroids with moderate aspect ratios would more efficiently marginate and adhere to the wall, as compared to spherical particles and prolate spheroids. Beyond size and shape, the effect of particle deformability on margination dynamics was documented in the work of Muller et al. (2016). These authors confirmed that micrometer carriers marginate better than their sub-micrometer counterparts and that deformable carriers are less prone to marginate as compared to rigid particles and demonstrated, in complex whole blood flow, that $2 \mathrm{D}$ and 3D simulations of red blood cells tend to return qualitatively similar results. Fish M. B. and colleagues explored the ability of micro- and nanosized particles to identify and bind to diseased endothelium, confirming that microcarriers outperforms nanoconstructs in margination and adhesive properties (Fish et al., 2017). In particular, rigid particles show improved adhesive abilities for large shear rate, while soft particles for low shear rates. 3D blood flow modeling and the accurate description of the viscoelastic properties of red blood cells (RBCs) has allowed the scientific community at large to learn more about molecular, nano and microscale transport within capillary networks, under physiological and pathological conditions (Fedosov et al., 2010; Ahmed et al., 2018; Li et al., 2017). Nonetheless, 2D models can still be very effective in dissecting some basic mechanisms regulating the interaction between deformable RBCs and nano/micro-particles. For instance, a recently published paper used a 2D blood flow model to predict the dispersion of nanoparticles, as a function of the RBC motion and deformation, in good agreement with experimental data (Tan et al., 2016). Despite these seminal works, at authors' knowledge, no study has ever systematically analyzed the interaction between particles deposited onto a wall and circulating blood cells.

Carrier margination and adhesion are necessary but not sufficient conditions for fully realizing drug delivery through vascular targeting. Indeed, carriers should adhere to the vessel walls for sufficiently long times, without being dislodged away by hemodynamic forces and circulating cells, in order to support the continuous and controlled release of therapeutic agents towards the diseased tissue. In this work, the authors propose a hybrid computational scheme, combining an ImmersedBoundary (IBM) and a Lattice Boltzmann-BGK (LBM) method, for studying cell-particle interactions in narrow capillaries. While the IBM serves to characterize the capillary transport of deformable objects (cells and particles), the LBM provides an Eulerian description of the fluid evolution (Coclite et al., 2016). While flowing within a narrow capillary, $7 \mu \mathrm{m}$ deformable cells encounter particles adhering to the wall, which act as partial occlusions. Cell and particle dynamics are predicted under different conditions, including four cell shapes, resembling leukocytes and erythrocytes; four aspect ratios for the adhering particles; two values of particle affinity with the vascular walls and two different values of the mechanical stiffness of the particles. Three parameters are introduced for quantitatively described the physical problem, namely the stretching ratio of the cell; relative distance between particles and cells; and adhesive strength of the particles to the wall.

\section{Computational method}
\subsection{The lattice Boltzmann method}
The evolution of the fluid is defined in terms of a set of $\mathrm{N}$ discrete distribution functions $\left\{f_{i}\right\},(i=0, \ldots, N-1)$, which obey the dimensionless Boltzmann equation
$$
f_{i}\left(\boldsymbol{x}+\boldsymbol{e}_{\boldsymbol{i}} \Delta t, t+\Delta t\right)-f_{i}(\boldsymbol{x}, t)=-\frac{\Delta t}{\tau}\left[f_{i}(\boldsymbol{x}, t)-f_{i}^{e q}(\boldsymbol{x}, t)\right],
$$
in which $\mathbf{x}$ and $t$ are the spatial and time coordinates, respectively; $\left[\boldsymbol{e}_{i}\right],(i=0, \ldots, N-1)$ is the set of discrete velocities; $\Delta \mathrm{t}$ is the time step; and $\tau$ is the relaxation time given by the unique non-null eigenvalue of the collision term in the BGK approximation (Bhatnagar et al., 1954). The kinematic viscosity of the flow is strictly related to the single relaxation time $\tau$ as $v=c_{s}^{2}\left(\tau-\frac{1}{2}\right) \Delta t$ being $c_{s}=\frac{1}{\sqrt{3}} \frac{\Delta x}{\Delta t}$ the reticular speed of sound. The moments of the distribution functions define the fluid density $\rho=\sum_{i} f_{i}$, velocity $\boldsymbol{u}=\sum_{i} f_{i} \boldsymbol{e}_{\boldsymbol{i}} / \rho$, and the pressure $p=c_{s}^{2} \rho=c_{s}^{2} \sum_{i} f_{i}$. The local equilibrium density functions $\left[f_{i}^{e q}\right](i=0, \ldots, N-1)$ are expressed by the Maxwell-Boltzmann distribution
$$
f_{i}^{e q}(\boldsymbol{x}, t)=\omega_{i} \rho\left[1+\frac{1}{c_{s}^{2}}\left(\boldsymbol{e}_{\boldsymbol{i}} \cdot \boldsymbol{u}\right)+\frac{1}{2 c_{s}^{4}}\left(\boldsymbol{e}_{\boldsymbol{i}} \cdot \boldsymbol{u}\right)^{2}-\frac{1}{2 c_{s}^{2}} \boldsymbol{u}^{2}\right] .
$$
On the two-dimensional square lattice with $\mathrm{N}=9$ speeds (D2Q9) (Qian et al., 1992), the set of discrete velocities is given by
$$
\boldsymbol{e}_{\boldsymbol{i}}= \begin{cases}(0,0), & \text { if } i=0 \\ \left(\cos \left(\frac{(i-1) \pi}{2}\right), \sin \left(\frac{(i-1) \pi}{2}\right)\right), & \text { if } i=1-4 \\ \sqrt{2}\left(\cos \left(\frac{(2 i-9) \pi}{4}\right), \sin \left(\frac{(2 i-9) \pi}{4}\right)\right), & \text { if } i=5-8,\end{cases}
$$
with the weight, $\omega_{\mathrm{i}}=1 / 9$ for $\mathrm{i}=1-4, \omega_{\mathrm{i}}=1 / 36$ for $\mathrm{i}=5-8$, and $\omega_{0}=4 / 9$. Here, we adopt a discretization in the velocity space of the equilibrium distribution based on the Hermite polynomial expansion of this distribution (Shan et al., 2006).

\subsection{Immersed-boundary treatment}
Deformable body models are commonly based on continuum approaches using strain energy functions to compute the membrane response (Pozrikidis, 2001; Skalak et al., 1973; Krüger, 2012). However, a particle-based model governed by molecular dynamics has emerged due to its mathematical simplicity while providing consistent predictions (Dao et al., 2006; Fedosov et al., 2011; Nakamura et al., 2013; Ye et al., 2014). Following this, here a particle-based model is adopted via the Immersed-Boundary (IB) technique. The immersed body consists in a network of $n v$ vertices linked with $n l$ linear elements, whose centroids are usually referred as Lagrangian markers. An effective forcing term $\mathcal{F}_{i}(i=0, \ldots, 8)$, accounting for the immersed boundary, is included as an additional contribution on the right-hand side of Eq. (1). $\mathcal{F}_{i}$ is expanded in term of the reticular Mach number, $\frac{e_{i}}{c_{s}}$, resulting in
$$
\mathcal{F}_{i}=\left(1-\frac{1}{2 \tau}\right) \omega_{i}\left[\frac{\boldsymbol{e}_{i}-\boldsymbol{u}}{c_{s}^{2}}+\frac{\boldsymbol{e}_{i} \cdot \boldsymbol{u}}{c_{s}^{4}} \boldsymbol{e}_{i}\right] \cdot \boldsymbol{f}_{i b},
$$
where $\boldsymbol{f}_{i b}$ is a body force term. Due to the presence of the forcing term, the macroscopic quantities are derived as
$$
\begin{aligned}
&\rho=\sum_{i} f_{i}, \\
&\rho \boldsymbol{u}=\sum_{i} f_{i} \boldsymbol{e}_{i}+\frac{\Delta \boldsymbol{t}}{2} \mathcal{F}_{i} .
\end{aligned}
$$
Notably, in such a framework, the forced Navier-Stokes equations can be recovered with a second order accuracy (Guo et al., 2002; De Rosis et al., 2014b, a; Suzuki et al., 2015; Wang et al., 2015). $\mathcal{F}_{i}$ accounts for the presence of an arbitrary shaped body immersed into the flow field, whereas the external boundaries of the computational domain are treated with the known velocity bounce back conditions by Zou and He (1997). The IBM procedure, which was proposed and validated by Coclite et al. (2016), is here adopted. The moving-least squares reconstruction by Vanella and Balaras (2009) is employed to exchange all LBM distribution functions between the Eulerian lattice and the Lagrangian chain, while the body force term in Eq. (4), $\boldsymbol{f}_{i b}$, is evaluated through the formulation of Favier et al. (2014).

\subsection{Elastic membrane deformation}
Membranes are subject to elastic strain response, bending resistance, and total enclosed area conservation. The stretching elastic potential acting on the two vertices sharing the lth element is given as
$$
V_{l}^{s}=\frac{1}{2} k_{s}\left(l_{l}-l_{l, 0}\right)^{2},
$$
being $k_{s}$ the characteristics elastic constant, $l_{l}$ the current length of the lth element, and $l_{l, 0}$ the length of the lth element in the stress-free configuration. Taking the derivative of the potential energy with respect to the displacements, the nodal forces corresponding to the elastic energy for nodes 1 and 2 connected by edge $l$ reads as
$$
\left\{\begin{array}{l}
\boldsymbol{F}_{1}^{s}=-k_{s}\left(l-l_{0}\right) \frac{\boldsymbol{r}_{1,2}}{l} \\
\boldsymbol{F}_{2}^{s}=-k_{s}\left(l-l_{0}\right) \frac{\boldsymbol{r}_{2,1}}{l},
\end{array}\right.
$$
where $\boldsymbol{r}_{i, j}=\boldsymbol{r}_{i}-\boldsymbol{r}_{j}$ with $\boldsymbol{r}_{i}$ position vector of the node $i$.

The bending resistance related to the $v$ th vertex connecting two adjacent element is
$$
V_{v}^{b}=\frac{1}{2} k_{b}\left(k_{v}-k_{v, 0}\right)^{2},
$$
being $k_{b}$ the characteristics bending constant, $k_{v}$ the current local curvature in the $v$ th vertex, $k_{v, 0}$ the local curvature in the $v$ th vertex for the stress-free configuration, which is evaluated by measuring the variation of the angle between the two adjacent elements $\left(\theta-\theta_{0}\right)$, with $\theta_{0}$ the angle in the stress free configuration. The forces on the nodes $v_{\text {left }}, v$, and $v_{\text {right }}$ are obtained considering the relation between the local curvature and the angle as
$$
\left\{\begin{array}{l}
\boldsymbol{F}_{v_{\text {left }}^{b}}^{b}=k_{b}\left(\theta-\theta_{0}\right) \frac{l_{\text {left }}}{l_{\text {left }} t l_{\text {right }}} \boldsymbol{n}_{v} \\
\boldsymbol{F}_{v}^{b}=-k_{b}\left(\theta-\theta_{0}\right) \boldsymbol{n}_{v} \\
\boldsymbol{F}_{v_{r i g h t}}^{b}=k_{b}\left(\theta-\theta_{0}\right) \frac{l_{\text {right }}}{l_{\text {left }} t l_{\text {right }}} \boldsymbol{n}_{v},
\end{array}\right.
$$
where $l_{\text {right }}$ and $l_{\text {left }}$ are the length of the two adjacent left and right edges, respectively, and $\boldsymbol{n}_{v}$ is the outward unity vector centered in $v$. In this simplified context for the bending resistance the relation between the strain response constant $k_{s}$ and $k_{b}$ is expressed through their ratio $\beta=\frac{k_{b}}{k_{s}}$.

In order to limit the membrane stretching, an effective pressure force term is considered. Thus, the penalty force is expressed in term of the reference pressure $p_{\text {ref }}$ and directed along the normal inward unity vector of the lth element $\boldsymbol{n}_{l}^{-}$, as
$$
\boldsymbol{F}_{l}^{a}=-k_{a}\left(1-\frac{A}{A_{0}}\right) p_{\text {ref }} \boldsymbol{n}_{l}^{-} l_{l},
$$
with $l_{l}$ the length of the selected element, $k_{a}$ the incompressibility coefficient, $A$ the current enclosed area, $A_{0}$ the enclosed area in the stress-free configuration. The enclosed area is computed using the Green's theorem along the curve, $A=\int x_{l} d y_{l}-$ $y_{l} d x_{l} . \boldsymbol{F}_{l}^{a}$ is equally distributed between the two vertices connecting the lth element $\left(v_{\text {left }}\right.$ and $\left.v_{\text {right }}\right)$ as $\boldsymbol{F}_{v_{\text {left }}}^{a}=\boldsymbol{F}_{v_{r i g h t}}^{a}=0.5 \boldsymbol{F}_{l}^{a}$. Note that, being $k_{a}$ a penalty function, its value should be selected as large as possible. For low Reynolds number flows, $k_{a}=1$ would return an incompressible membrane, as demonstrated in Supp. Figure.1 for the circular and bullet shaped cells and in Supp.Figure.2 for the biconcave and parachute shaped cells. Moreover, a deformed cell would restore its initial configuration, see Supp.Figure.3, for null external stresses (reproducing quiescent conditions with zero hydrodynamics stresses) and the cell continues to evolve uniquely under the effect of internal stresses.

\subsection{Particle-particle interaction modeling}
Two-body interactions are modeled through a repulsive potential centered in each vertex composing the immersed particles. The repulsive force is such that the minimum allowed distance between two vertices coming from two different particles is $\Delta x$. The impulse acting on vertex 1 , at a distance $d_{1,2}$ from the vertex 2 of an adjacent particle, is directed in the inward normal direction identified by $\boldsymbol{n}_{1}^{-}$and is given by
$$
\boldsymbol{F}_{1}^{p p}=\frac{10^{-4}}{8 \sqrt{2}} \sqrt{\frac{\Delta x}{d_{1,2}^{5}}} \boldsymbol{n}_{1}^{-} .
$$

\subsection{Wall-particle interaction}
Ligand and receptor molecules are distributed over the surfaces of the particle and blood vessel wall (Coclite et al., 2017; Sun et al., 2003). Ligand molecules are modeled as linear springs, which tend to establish bonds with receptors on the vascular wall, resulting in a mechanical force given as
$$
\boldsymbol{F}_{l}^{w p}=\sigma\left(y_{l}-y_{c r, e q}\right) \boldsymbol{n}_{l},
$$
with $y_{l}$ the bond length, $y_{c r, e q}$ the equilibrium bond length and $\sigma$ the spring constant (same for all springs). Bonds can be only generated if the minimum separation distance between the particle boundary and the wall is smaller than a critical value, $y_{c r}=25 \mathrm{~nm}$ (Roy and Qi, 2010) whereas the equilibrium bond length, resulting in a null force, is $y_{c r, e q}=0.5 y_{c r}$. The linear spring constant is computed in lattice units through its dimensionless group, $\frac{\rho_{\text {ref }} v_{\text {ref }}^{2}}{H}$, where $\rho_{\text {ref }}, H$, and $v_{\text {ref }}$ are the reference density, length and viscosity, respectively. At each time step, the bond formation is regulated by a forward probability function, while a reverse probability function controls the destruction of a pre-existing bond, see for details (Coclite et al., 2017). The adhesive force, being calculated at the centroid of each element, is evenly distributed to the two vertices connecting the element in the same fashion used for $\boldsymbol{F}^{a}$.

\subsection{Hydrodynamics stresses}
Pressure and viscous stresses exerted by the lth linear element are
$$
\begin{aligned}
&\boldsymbol{F}_{l}^{p}(t)=\left(-p_{l} \boldsymbol{n}_{l}\right) l_{l}, \\
&\boldsymbol{F}_{l}^{\tau}(t)=\left(\bar{\tau}_{l} \cdot \boldsymbol{n}_{l}\right) l_{l},
\end{aligned}
$$
where $\bar{\tau}_{l}$ and $p_{l}$ are the viscous stress tensor and the pressure evaluated in the centroid of the element, respectively; $\boldsymbol{n}_{l}$ is the outward normal unit vector while $l_{l}$ is its length. The pressure and velocity derivatives in Eqs. (14) and (15) are evaluated considering a probe in the normal positive direction of each element, being the probe length $1.2 \Delta x$, and using the cited moving least squares formulation (Coclite et al., 2016). In this framework, the velocity derivatives evaluated at the probe are considered equal to the ones on the linear element centroid as previously done by the authors (Coclite et al., 2016; de Tullio and Pascazio, 2016). In this case, all force contributions are computed with respect to the centroid of each elements and then transferred to the vertices.

\subsection{Fluid-structure interaction}
The motion of the immersed bodies is described through the solution of the Newton equation for each Lagrangian vertex, accounting for both internal, Eqs. (8), (10), and (11), and external stresses, Eqs. (12), (13), (14), and (15).

\subsection{Deformable structures motion}
The total force $\boldsymbol{F}_{v}^{\text {tot }}(t)$ acting on the $v$ th element of the immersed body is evaluated in time and the position of the vertices is updated at each Newtonian dynamics time step considering the membrane mass uniformly distributed over the $n v$ vertices (Buxton et al., 2005; Ahlrichs and Dunweg, 1998) as
$$
m_{v} \dot{\boldsymbol{u}}_{v}=\boldsymbol{F}_{v}^{t o t}(t)=\boldsymbol{F}_{v}^{s}(t)+\boldsymbol{F}_{v}^{b}(t)+\boldsymbol{F}_{v}^{a}(t)+\boldsymbol{F}_{v}^{p p}(t)+\boldsymbol{F}_{v}^{w p}(t)+\boldsymbol{F}_{v}^{p}(t)+\boldsymbol{F}_{v}^{\tau}(t) .
$$
The Verlet algorithm is employed to integrate the Newton equation of motion. Precisely, a first tentative velocity is considered into the integration process, $\dot{\boldsymbol{x}}_{v, 0}(t)$, obtained interpolating the fluid velocity from the surrounding lattice nodes
$$
\boldsymbol{x}_{v}(t+\Delta t)=\boldsymbol{x}_{v}(t)+\dot{\boldsymbol{x}}_{v, 0}(t) \Delta t+\frac{1}{2} \frac{\boldsymbol{F}_{v}^{t o t}(t)}{m_{v}} \Delta t^{2}+O\left(\Delta t^{3}\right) .
$$
Then, the velocity at time level $t+\Delta t$ is computed as
$$
\boldsymbol{u}_{v}(t+\Delta t)=\frac{\frac{3}{2} \boldsymbol{x}_{v}(t+\Delta t)-2 \boldsymbol{x}_{v}(t)+\frac{1}{2} \boldsymbol{x}_{v}(t-\Delta t)}{\Delta t}+O\left(\Delta t^{2}\right) .
$$

\subsection{Rigid structures motion}
Rigid motion is readily obtained integrating all stress contributions over the particles boundary and updating both linear and angular velocity in time as $\dot{\boldsymbol{u}}(t)=\frac{\boldsymbol{F}^{\text {tot }}(t)}{m}$ and $\dot{\omega}(t)=\frac{M^{\text {tot }}(t)}{I}$; where $m$ is the particle mass; $\boldsymbol{F}^{\text {tot }}(t)$ is the total force exerted by the particle; $M^{\text {tot }}(t)$ is the total moment acting on the particle; and $I$ is the moment of inertia. Lastly, $\boldsymbol{u}(t)$ and $\omega(t)$ are computed as
$$
\begin{aligned}
&\boldsymbol{u}(t)=\frac{2}{3}\left(2 \boldsymbol{u}(t-\Delta t)-\frac{1}{2} \boldsymbol{u}(t-2 \Delta t)+\dot{\boldsymbol{u}}(t) \Delta t\right)+O\left(\Delta t^{2}\right) \\
&\omega(t)=\frac{2}{3}\left(2 \omega(t-\Delta t)-\frac{1}{2} \omega(t-2 \Delta t)+\dot{\omega}(t) \Delta t\right)+O\left(\Delta t^{2}\right)
\end{aligned}
$$
with $\Delta x=\Delta t=1$.

The proposed approach is unconditionally stable for small local velocity variations within any solid/fluid density ratio (Zhang and Hisada, 2004), which is indeed the case of the present work.

\section{Results and Discussion}
\subsection{Geometry and boundary conditions of the problem}
The combined Lattice Boltzmann-Immersed Boundary method, described in the previous section, is here used to predict the interaction dynamics of deformable cells with particles, which are originally sitting on the walls of a narrow vessel (see Table 1 and Fig. 1). A capillary flow is modeled within a rectangular channel with a height of $7.5 \mu \mathrm{m}$ and a length of $75 \mu \mathrm{m}$. At the inlet section of the channel $(x=0)$, a parabolic velocity profile is imposed; at the outlet section of the channel $(x=10 \mathrm{H})$, a constant pressure condition is applied $\left(p=p_{\text {ref }}\right)$; and, on the walls of the channel, a no-slip velocity condition is imposed. This results in a plane-Poiseuille flow along the channel, which moves blood cells downstream and challenge the vascular adhesion of the particles. Given a kinematic viscosity $v=1.2 \times 10^{-6} \mathrm{~m}^{2} / \mathrm{s}$; a center line velocity $u_{\max }=1.6 \times 10^{-3} \mathrm{~m} / \mathrm{s}$, the Reynolds number is $\operatorname{Re}=u_{\max } \mathrm{H} / v=10^{-2}$.

At time zero, a deformable blood cell is released, with zero velocity, at the capillary center line $\left(x_{C_{0}}=5.0 \mu \mathrm{m} ; y_{C_{0}}=\right.$ $3.75 \mu \mathrm{m})$ and transported downstream by the flow. The cell has the same density as the surrounding fluid $\left(\rho=10^{3}\right.$ $\mathrm{a}$ 
$\mathrm{kg} / \mathrm{m}^{3}$ ) and is slightly smaller than the height of the channel. The cell deformability is quantified via the Capillary number $C a=10^{-3}\left(=\rho v u_{\max } / k_{s}\right)$ and null bending resistance (Pozrikidis, 2005; Omori et al., 2012). In particular, the stress-free biconcave shape is obtained as the parametrization of the median section of a red blood cell (Pozrikidis, 2003; Sui et al.,2007),
$$
\left\{\begin{array}{l}
x=a \alpha \sin q \\
y=a \frac{\alpha}{2}\left(0.207+2.003 \sin ^{2} q-1.123 \sin ^{4} q\right) \cos q,
\end{array}\right.
$$
where $a$ is $0.122, \alpha$ is the cell radius, $3.5 \mu \mathrm{m}$, and the parameter $q$ varies in $[-0.5 \pi, 1.5 \pi]$. Within this parametrization the swelling ratio of the stress free configuration correctly corresponds to $0.481$ (Shi et al., 2012). Indeed, the present parametrization deals with the deformation of a blood cell in a rectangular 2D channel in which a plane Hagen-Poiseuille flow is established. This may differ from the deformation of a RBC in a 3D circular capillary, being the deformation of immersed structures strongly dependent on the hydrodynamics stresses, and the latter being different between a $2 \mathrm{D}$ rectangular channel and a 3D circular capillary (Ahmed et al., 2018; Li et al., 2017; Tan et al., 2016; Fedosov et al., 2010).
\begin{figure}
\centering
\includegraphics[scale=0.35]{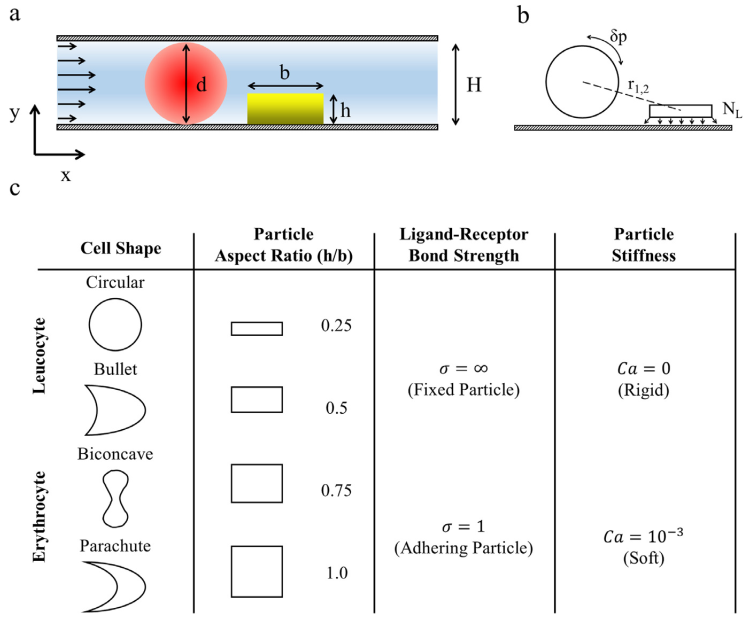}
\caption{Schematic of the physical problem. a. Schematic representation of the computational domain with characteristic dimensions. b. Sketch of the computed physical quantities. c. Cells and particles characterization in terms of membrane geometry, dimension, adhesive properties and mechanical stiffness.}
\label{Fig1}
\end{figure}
\begin{figure}
{\small {\bf Table.1} Parameters used in the computational experiments expressed in the SI-unit system and in lattice-unit system along with their dimensionless groups. Note that, the channel thickness, $H$; the water density, $\rho_{\text {ref }}$; the water kinematic viscosity, $v_{\text {ref }}$ are used throughout the formulation to normalize all dependent variables.}

\includegraphics[scale=0.45]{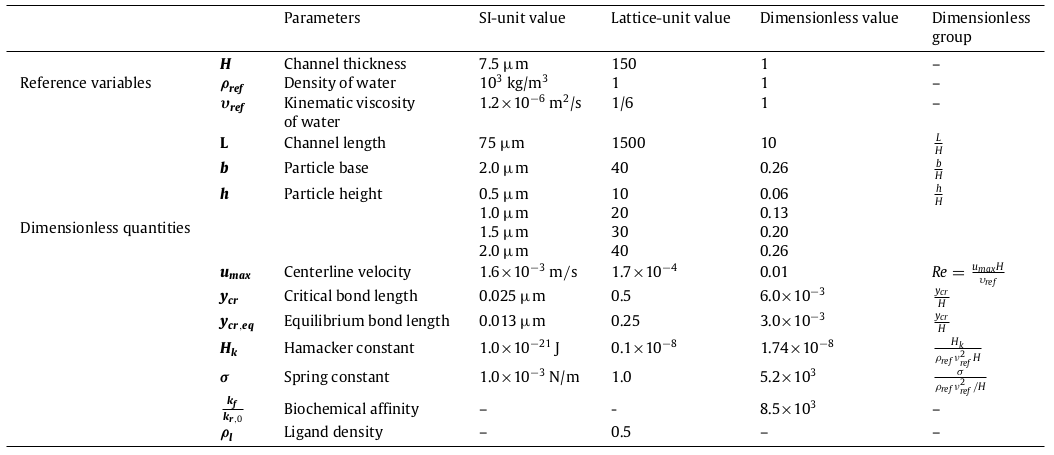}
\label{Table1}
\end{figure}
On the lower wall of the channel, a rectangular particle is deposited with a height $h$ ranging between $0.5$ and $2.0 \mu \mathrm{m}$ and a fixed length $b=2.0 \mu \mathrm{m}$. The particle is located next to the wall $\left(x_{P_{0}}=12.5 \mu \mathrm{m} ; y_{P, o}=12.5 \mathrm{~nm}\right)$, returning an initial separation distance from the cell $r_{o}=1.6 \mathrm{H}$. The particle surface is decorated by ligand molecules with a density $\rho_{\mathrm{L}}=0.5$. These ligands interact with receptors expressed on the capillary walls forming ligand-receptor bonds with a biochemical affinity given by $k_{\mathrm{f}} / k_{\mathrm{r}, 0}=8.5 \times 10^{3}$ (Sun et al., 2003; Sun and Munn, 2008; Chang and Hammer, 1996; Chang et al., 2000). The density of the particle is slightly higher than that of the surrounding fluid, being $\rho_{P} / \rho=1.1$. Following the findings of Mody and King (2007), thermal fluctuations are neglected in the present work, being almost uninfluential for micrometric and sub-micrometric particles under flow.

In order to assess the adhesive strength of the particle to the wall, three physical parameters are monitored over time (Fig. 1b): the cell stretching ratio $\delta p\left(=\frac{p-p_{0}}{p_{0}}\right)$, which is defined as the relative variation in cell perimeter $p$ over the initial configuration $p_{o}$; the relative cell-to-particle separation distance $r\left(=\sqrt{\left(x_{C}-x_{P}\right)^{2}+\left(y_{C}-y_{P}\right)^{2}}\right)$, which is defined as the absolute distance between the centroids of the cell and the particle, being $x_{C}\left(x_{P}\right)$ and $y_{C}\left(y_{P}\right)$ the coordinates of the cell (particle) centroids; and the number $N_{L}$ of ligand-receptor bonds engaged at the interface between the particle and capillary wall. As summarized in the table of Fig. 1c, the temporal evolution of these three physical quantities is systematically analyzed by varying four independent parameters: the cell shape-circular and bullet-shaped cells, mimicking leukocytes; biconcave and parachute-shaped cells, mimicking erythrocytes; the particle aspect ratio $h / b-0.25,0.5,0.75$ and $1.0$; the strength of the ligand-receptor bonds $-\sigma=1$ and $\infty$ (infinitely strong bond); and the particle deformability - Ca $=10^{-3}$ and 0 (rigid particle). Note that, bullet- and parachute-shaped cells are transported considering as stress-free configurations the circular and the biconcave membranes, respectively. The bullet- and parachute-shaped cells, indeed, are the initially circular and biconcave membranes that have reached the final deformed configuration in the plane Hagen-Poiseuille flow investigated in this work (see Supp.Figure.4). This approach allows, in turn, a thorough analysis (in term of cell shapes) of the dynamics behavior of the presented cell-particle system.

\subsection{Cell-particle interaction in narrow capillaries: a qualitative description}
\begin{figure}
\centering
\includegraphics[scale=0.5]{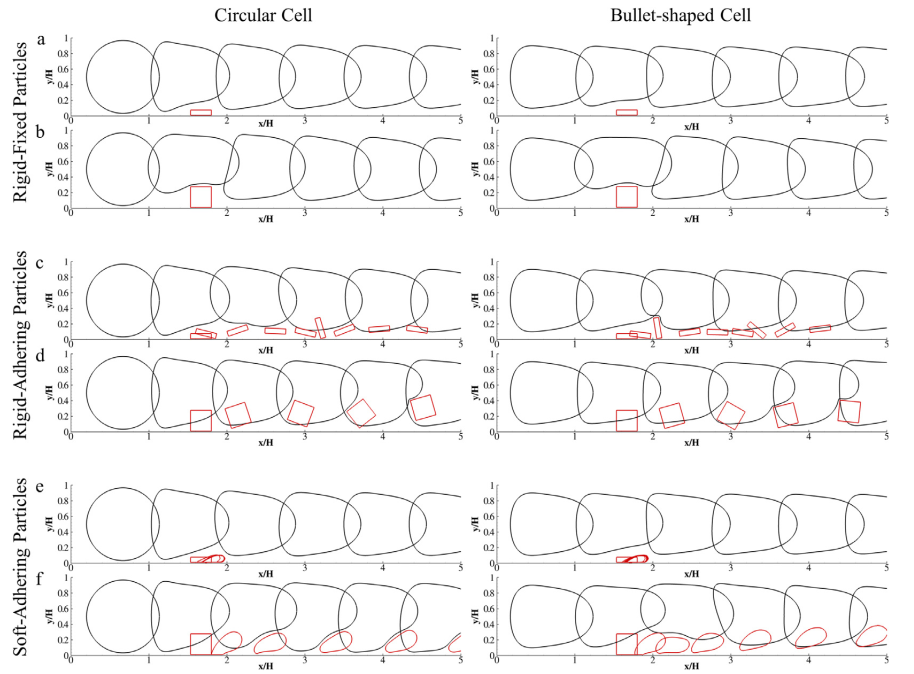}
\caption{Dynamics of circular and bullet shaped cell over a particle deposited on the wall. Cell and particle configurations, taken at six different time points $t \frac{u_{\max }}{H}=0,1,2,3,4$, and 5 , along the capillary. The particle is rigid and fixed to the wall $(\mathrm{a}, \mathrm{b})$; rigid and adhering to the wall (c,d); soft and adhering to the wall (e,f). The particle aspect ratios are $\mathrm{h} / \mathrm{b}=0.25$ (upper lines) and $1.0$ (lower lines) for each considered cases.}
\label{Fig2}
\end{figure}
\begin{figure}
\centering
\includegraphics[scale=0.5]{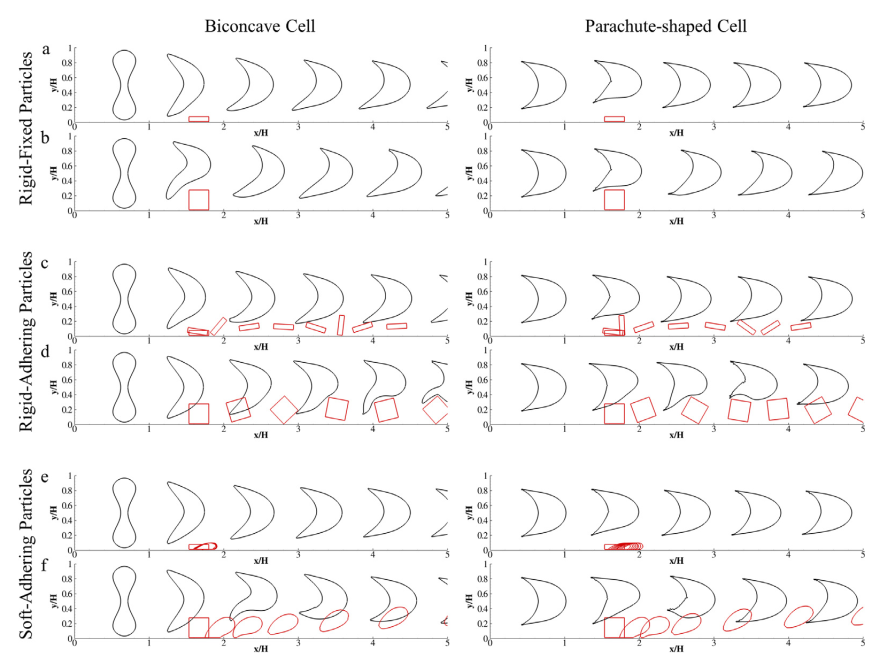}
\caption{Dynamics of biconcave and parachute shaped cell over a particle deposited on the wall. Cell and particle configurations, taken at six different time points $t \frac{u_{\max }}{H}=0,1,2,3,4$,and 5 , along the capillary. The particle is rigid and fixed to the wall (a,b); rigid and adhering to the wall (c,d); soft and adhering to the wall (e,f). The particle aspect ratios are $\mathrm{h} / \mathrm{b}=0.25$ (upper lines) and $1.0$ (lower lines) for each considered cases.}
\end{figure}
A circular, deformable cell is used to mimic a leucocyte moving along the capillary. Fig. 2 shows a series of snapshots from simulations documenting the interaction of such a cell with a particle, originally sitting on the lower wall of the capillary. Three different conditions have been tested, namely a rigid particle fixed on the wall (Fig. $2 \mathrm{a}, \mathrm{b})(\mathrm{Ca}=0 ; \sigma=\infty) ;$ a rigid particle adhering on the wall (Fig. $2 \mathrm{c}, \mathrm{d})(\mathrm{Ca}=0 ; \sigma=1)$; and a soft particle adhering on the wall $($ Fig. $2 \mathrm{e}, \mathrm{f})\left(\mathrm{Ca}=10^{-3} ; \sigma=1\right)$. For each condition, the upper line corresponds to a thin particle (aspect ratio: $0.25$ ) while the lower line corresponds to a thick particle (aspect ratio: $1.0$ ). Along the channel, cell deformation is dictated by the complex hydrodynamic conditions induced by the Poiseuille flow as well as by the presence of the particle on the wall. The latter represents an obstacle along the way of the cell that locally reduces the caliber of the vessel. The higher is the particle aspect ratio (h/b) and the higher is the obstruction and, consequently, the higher is the deformation experienced by the cell. This is clearly shown in Fig. 2a,b for thin (h/b $=0.25$ ) a and thick (h/b $=1.0)$ rigid particle firmly adhering on the wall $(\sigma=\infty)$. For thin particles, a modest deviation of the cell shape from the equilibrium configuration (undisturbed capillary flow - Cell Alone) is observed, whereas for thicker particles, a more significant deformation of the cell membrane is documented. This is also presented in the Supplementary Movies. 1 and 2. Under the hypothesis of a rigid adhering $(\sigma=1)$ rather than fixed particle, both thin and thick particles are dislodged away and carried along by the cell, as depicted in Fig. 2c, and d. (see also Supplementary Movie.3 and 4 ) The rigid particles roll over the wall following a complex dynamics, which is governed by their geometry and the interaction with the cell. Finally, in the case of soft particles adhering to the wall, see Fig. $2 \mathrm{e}, \mathrm{f},\left(\mathrm{Ca}=10^{-3} ; \sigma=1\right)$, both the particle and the cell deform. For thin soft particles, deformation adsorbs part of the energy provided by the cell thus preventing abrupt detachment (Fig. 2 e and Supplementary Movie.5). Differently, thicker particles are deformed and dislodged away by the moving cell (Fig. $2 \mathrm{f}$ and Supplementary Movie.6). Similar behaviors are depicted for all tested cell shapes, as demonstrated in the right column of Fig. 2 for bullet shaped cells and in Fig. 3 for erythrocytes. Here, the adhesive ability of squared particles is compared with that of circular particles, exhibiting a diameter $\mathrm{d}=0.5,1.0,1.5$, and $2.0 \mu \mathrm{m}$. Note that, this diameter is identical to the thickness of the square particles thus providing similar blockage ratios. As compared to the square particles, the spherical particles and the limited size of their contact area support continuous rolling over the vessel walls rather than adhesion (Supplementary Movie.7 and .8). Small soft particles establish a steady rolling motion (Supplementary Movie.9) while, soft d $=2.0 \mu \mathrm{m}$ particles are dislodged away from the wall (Supplementary Movie.10). These trends are confirmed by experimental data obtained by Charoenphol and colleagues (Charoenphol et al., 2012). In fact, they demonstrate that leukocytes prevent the formation of stable chemical bonds between microspherical particles and the endothelium in laminar flows.

\subsection{Cell-particle interaction in narrow capillaries}
Cell-particle interactions are more accurately described by quantifying the variation of the three dependent physical parameters - $\delta \mathrm{p}, \mathrm{r}, N_{\mathrm{L}}$ - over time. In the case of a circular, deformable cell moving on a firmly adhering particle (Ca $=0$, $\sigma=\infty$ ), the cell stretching, $\delta$ p, increases significantly as long as the cell is in close proximity to the particle (Fig. 4a). As expected, the higher is the particle aspect ratio and the higher is the cell deformation. Note that the flow per se is already deforming the cell so that $\delta$ p tends asymptotically to $0.034$ away from the particle (equilibrium cell shape under flow - Cell Alone). However, in the presence of a thick particle $(\mathrm{h} / \mathrm{b}=1)$, the cell membrane deformation returns a maximum $\delta \mathrm{p}$ as high as $0.088$. The maximum relative stretching is achieved when the cell and particle are very close to each other, or in other words, when $r$ reaches a minimum, as shown in Fig. 4b. In the case of rigid particles adhering on the wall $(\mathrm{Ca}=0, \sigma=1)$, the cell deformation follows a more complex pattern with two local maxima (Fig. 4c). The first maximum is associated with the initial cell-particle interaction that eventually results in the detachment and downstream transport of the particle. The second maximum are related to the interaction of the cell with the dislodged particle, in the flow. This is confirmed by the variation in the cell-particle separation distance $r$ (Fig. 4d), which shows a wide and flat minimum documenting the fact that the particle is dragged away and stays in close proximity of the moving cell. This is also clearly presented in the snapshots of Figs. $2 \mathrm{c}, \mathrm{d}$ and $3 \mathrm{c}$, d, where the rolling particles appear to interact with the moving cells in the flow. Overall, this behavior is observed for all particles regardless of their aspect ratio. Thinner particles exhibit a larger primary peak followed by a more moderate secondary peak; whereas thicker particles present a modest primary peak followed by a larger secondary peak. Indeed, particle geometry affects the location and amplitude of the maxima and thinner particles are more easily passed by the faster moving cells thus explaining the moderate secondary peaks. In the case of deformable particles adhering on the wall ( $\mathrm{Ca}=10^{-3}, \sigma=1$ ), cell deformation is diminished and becomes significant only for thicker particles, which eventually are dislodged away and carried along by the moving cells (Fig. 4e,f). Overall, the cell-stretching $\delta$ p increases only up to $0.062$ for thicker particles (Fig. 4e). Interestingly, these particles, which are dislodged away from the moving cell, tend to interact for a longer time with the cell and present a quasi-constant separation distance $r$ (Figs. $2 \mathrm{f}$ and 3f). In summary, while thin, soft particles are squeezed down against the wall by the moving cell, thick, soft and all rigid particles are dislodged away by the flow and the moving cell. Similar results are obtained for different cell shapes, as documented in Supplementary Figure. 5 for the bullet shaped cells and in Supplementary Figure.6 and .7 for the erythrocytes.
\begin{figure}
\centering
\includegraphics[scale=0.5]{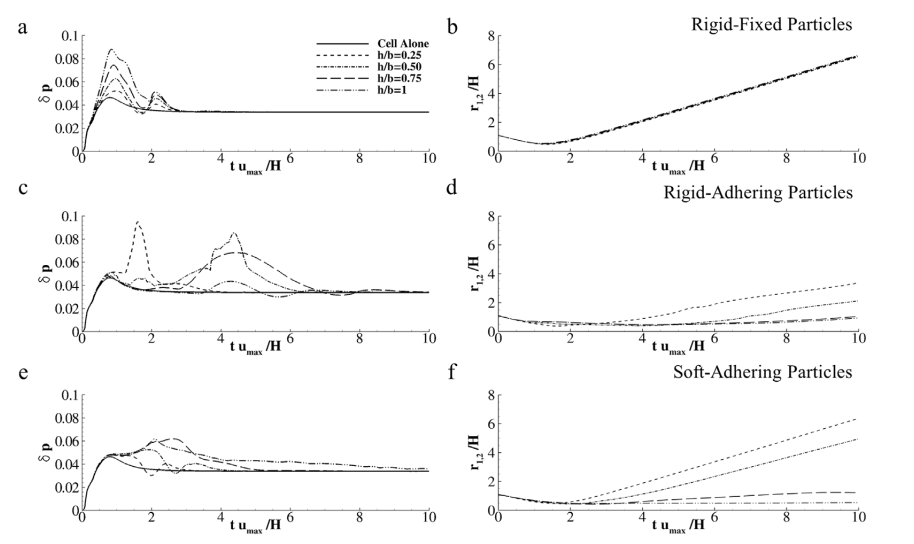}
\caption{Stretching ratio and separation distance for the transported circular cell. (a, c, e) Variation of the stretching ratio $\delta p$ for a circular cell moving over a particle with aspect ratios $\mathrm{h} / \mathrm{b}=0.25,0.5,0.75$, and 1.0. (b, d, f) Variation of the separation distance between the centroids of the blood cell and particle with aspect ratios $\mathrm{h} / \mathrm{b}=0.25,0.5,0.75$, and $1.0$. The Cell Alone case implies that the particle is not present in the capillary.}
\end{figure}
Finally, the number of active ligand-receptor bonds $N_{L}$ is presented in Fig. 5 for circular rigid $(\mathrm{Ca}=0 ; \sigma=1)$ and soft ( $\mathrm{Ca}=10^{-3} ; \sigma=1$ ) adhering particles. At time $\mathrm{t}=0$, all ligand-receptor bonds distributed over the base of the particles are engaged, thus $N_{L}=1$. For rigid particles, which are displaced away from the moving cell, $N_{L}$ rapidly decreases to zero within $t u_{\max } / H=2$ (Fig. 5, left column). This detachment phase is followed by multiple, but ephemeral, phases of transient, rapid ligand-receptor bond formation and disruption. This is only observed for the thinner particles whereby $N_{L}$ shows peaks up to $0.25$ and $0.5$ for an aspect ratio of $0.25$ and $0.5$, respectively. Differently, thicker particles, with an aspect ratio higher than $0.5$, are rapidly displaced away by the fluid flow and cell, and can only form a very modest number of bonds during their rolling (Fig. 5, left column). In these cases, $N_{L}$ does not grow higher than $0.025$, as shown in the insets of the same figures. The case with no cells (Particle Alone) is also presented in Fig. 5. Interestingly, multiple and ephemeral ligand-receptor bonds are formed in this case. Note that the peaks are similar in intensity but are slightly shifted, when comparing the cases with and without cells.
\begin{figure}
\centering
\includegraphics[scale=0.5]{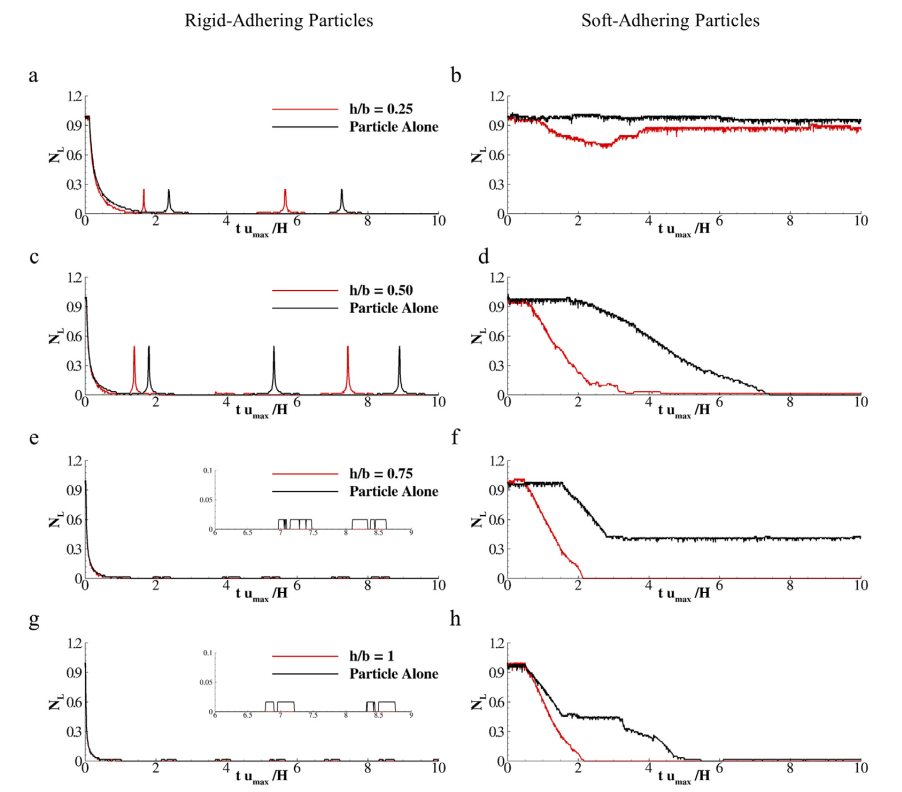}
\caption{Number of engaged bonds at the particle-wall interface. Number of engaged bonds over time for rigid (a, c, e, g) and soft particles (b, d, f, h). The case Particle Alone implies that no blood cells are transported along the capillary. ( $\mathrm{N}_{\mathrm{L}}$ is normalized with respect to the number of closed bonds in the initial configuration: $\left.\mathrm{N}_{\mathrm{L}, 0}=60.\right)$.}
\end{figure}
In the case of deformable particles ( $\mathrm{Ca}=10^{-3} ; \sigma=1$ ), depicted in Fig. 5 (right column), the pattern is more complex and depends on the particle geometry. As already noted, particle deformation enables the accumulation of a portion of the kinetic energy provided by the moving cell as elastic energy. Consequently, this reduces the mechanical force transferred to the ligand-receptor bonds and increases the overall strength of particle adhesion. For an aspect ratio of $0.25, N_{L}$ slightly reduces and then grows again recovering the value at equilibrium (Particle Alone). As the aspect ratio grows, the number of active ligand-receptor bonds decreases and goes to zero (detachment). The time for detachment decreases with the aspect ratio: $t u_{\max } / H=3.4$, for an aspect ratio of $0.5 ; t_{\max } / H=2.18$, for an aspect ratio of $0.75$ and 1.0. Interestingly, in the case of Particle Alone with an aspect ratio of $0.75, N_{\mathrm{L}}$ reduces and reaches a plateau for $N_{\mathrm{L}}=0.41$. This behavior has to be ascribed to particle deformation (see Supplementary Figure.8). As the area of adhesion to the wall reduces slightly at the trailing edge, the particle deforms and bends over the wall on its lateral side at the leading edge. This balances the loss of adhesion at the trailing edge and allows the particle to find a new configuration of equilibrium. This is not observed for thin particles, in that their side edge is smaller, and for thick particles, because of the larger dislodging hydrodynamic forces. Finally, the effect of the cell shape on the temporal variation of $N_{\mathrm{L}}$ is shown in Fig. 6 . In the case of a thin particle (h/b= $0.25$ ), only minor differences can be observed. In general, erythrocyte-like cells introduce smaller perturbations resulting in more stable bonds with $N_{\mathrm{L}}$ being very close to the original value of $1.0$. Similar general observations can be presented for thicker particles. A more complex behavior is observed for particles with intermediate aspect ratios (Fig. 6b, c). Biconcave and bullet shaped cells appear to be favoring the wall adhesion of particles with an aspect ratio of $0.5$. This should possibly be ascribed to an overall milder interaction of the cell with the particle, resulting in lower dislodging forces. The extremities of the biconcave and bullet shaped cells, which come closer to the particle surface, are more easily deformed as compared to the more bulky circular and parachute shaped cells. Similar observations can be applied to particles with an aspect ratio of $0.75$ exposed to bullet shaped cells. In this case, given the longer side edge, the particle is squeezed against the wall and adheres on its lateral edge at the leading edge, returning a $N_{\mathrm{L}}=1$.17. The deformation of particles with aspect ratios of $0.5$ and $0.75$ is depicted in the Supplementary Figure.9.
\begin{figure}
\centering
\includegraphics[scale=0.4]{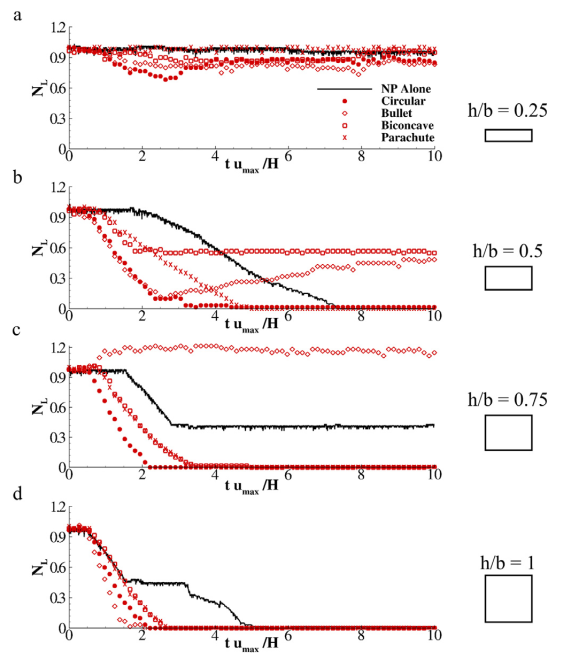}
\caption{Number of engaged bonds at the particle-wall interface. Number of engaged bonds over time for rigid (a, c, e, g) and soft particles (b, d, f, h). The case Particle Alone implies that no blood cells are transported along the capillary. ( $\mathrm{N}_{\mathrm{L}}$ is normalized with respect to the number of closed bonds in the initial configuration: $\left.\mathrm{N}_{\mathrm{L}, 0}=60.\right)$.)}
\end{figure}

\section{Conclusions and future work}
A combined Lattice Boltzmann -Immersed Boundary Method is employed to predict the interaction between a deformable cell and vascular targeted particles in narrow capillaries. Particle adhesion to the capillary wall is challenged by hemodynamic forces and flow perturbations induced by circulating blood cells. The strength of adhesion is systematically assessed as a function of the cell geometry - circular and bullet shaped cells for leukocytes; biconcave and parachute shaped cells for erythrocytes; particle aspect ratio - ranging from $0.25$ to $1.0$; adhesion affinity with the capillary wall - fixed particle and ligand-receptor bond mediated adhesion; and mechanical stiffness - rigid and soft.

It is observed that rigid particles of any aspect ratio would be immediately detached away from the wall and dislodged along the capillary by circulating blood cells. The detached particle would roll over the blood vessel walls establishing shortlived adhesive interactions but failing to achieve firm adhesion again. Differently, soft particles would deform and transform part of the kinetic energy provided by circulating cells into elastic energy, thus reducing the forces directly exerted on the ligand-receptor bonds. This increases the likelihood of deformable particles to comply with the flow perturbations induced by the circulating cells. Yet, only sufficiently thin, soft particles would resist full detachment from the blood vessel walls.

Collectively, these data demonstrate that firm vascular adhesion can be more efficiently achieved by using deformable carriers as compared to their rigid counterpart. Soft and sufficiently thin carriers would firmly adhere to the capillary walls allowing for the sustained release of therapeutic agents from the lumen towards the diseased tissue. Indeed, a 3D representation of the problem at hand would allow for a more precise characterization of the 'swept off' critical conditions (particle thickness and elastic modulus), without changing the final conclusions of the presented 2D model. In a future work, it would certainly be interesting to generate a phase diagram identifying the critical 'swept off' conditions in terms of the particle properties (size, shape, surface ligand density and mechanical stiffness) and local hydrodynamic conditions (wall shear rate), in a 3D model of blood flow.

\section{Acknowledgments}
This project was partially supported by the European Research Council, under the European Union's Seventh Framework Programme (FP7/2007-2013)/ERC grant agreement no. 616695; AIRC (Italian Association for Cancer Research) under the individual investigator grant no. 17664; European Union's Horizon 2020 research and innovation programme under the Marie Sadowska-Curie grant agreement no. 754490 “MINDED”; the Italian Institute of Technology.

\section{References}
Ahlrichs, P., Dunweg, B., 1998. Lattice-Boltzmann simulation of polymer-solvent systems. Internat. J. Modern Phys. C 9 (8), 1429-1438.

Ahmed, F., et al., 2018. Internal viscosity-dependent margination of red blood cells in microfluidic channels. J. Biomech. Eng. 140 (6), 061013.

Anselmo, A.C., et al., 2015. Elasticity of nanoparticles influences their blood circulation, phagocytosis, endocytosis, and targeting. ACS Nano 9, $3169-3177$.

Anselmo, A.C., Mitragotri, S., 2017. Impact of particle elasticity on particle-based drug delivery systems. Adv. Drug Delivery Rev. 108 (Supplement C), $51-67$.

Atukorale, P.U., et al., 2017. Vascular targeting of nanoparticles for molecular imaging of diseased endothelium. Adv. Drug Delivery Rev. 113, 141-156.

Bhatnagar, P.L., Gross, E.P., Krook, M., 1954. A model for collision processes in gases. I. Small amplitude processes in charged and neutral one-component systems. Phys. Rev. 94 (3), 511-525.

Buxton, G.A., et al., 2005. Newtonian fluid meets an elastic solid: coupling lattice Boltzmann and lattice-spring models. Phys. Rev. E (3) 71 (5 Pt 2), 056707.

Chang, K.-C., Hammer, D.A., 1996. Influence of direction and type of applied force on the detachment of macromolecularly-bound particles from surfaces. Langmuir 12 (9), 2271-2282.

Chang, K.-C., Tees, D.F., Hammer, D.A., 2000. The state diagram for cell adhesion under flow: leukocyte rolling and firm adhesion. Proc. Natl. Acad. Sci. $97(21), 11262-11267$

Charoenphol, P., et al., 2012. Particle-cell dynamics in human blood flow: implications for vascular-targeted drug delivery. J. Biomech. 45 (16), $2822-2828$.

Coclite, A., et al., 2016. A combined Lattice Boltzmann and Immersed boundary approach for predicting the vascular transport of differently shaped particles. Comput. \& Fluids 136, 260-271.

Coclite, A., et al., 2017. Predicting different adhesive regimens of circulating particles at blood capillary walls. Microfluid. Nanofluid. 21 (11), 168.

Dao, M., Li, J., Suresh, S., 2006. Molecularly based analysis of deformation of spectrin network and human erythrocyte. Mater. Sci. Eng. C-Biomimetic and Supramolecular Syst. 26 (8), 1232-1244.

Daquinag, A.C., Zhang, Y., Kolonin, M.G., 2011. Vascular targeting of adipose tissue as an anti-obesity approach. Trends Pharmacol. Sci. 32 (5), $300-307$.

De Rosis, A., Ubertini, S., Ubertini, F., 2014a. A comparison between the interpolated bounce-back scheme and the immersed boundary method to treat solid boundary conditions for laminar flows in the Lattice Boltzmann framework. J. Sci. Comput. 61 (3), 477-489.

De Rosis, A., Ubertini, S., Ubertini, F., 2014b. A partitioned approach for two-dimensional fluid-structure interaction problems by a coupled lattice Boltzmann-finite element method with immersed boundary. J. Fluids Struct. 45, 202-215.

Decuzzi, P., Ferrari, M., 2008. Design maps for nanoparticles targeting the diseased microvasculature. Biomaterials 29 (3), $377-384$.

Euliss, L.E., et al., 2006. Imparting size, shape, and composition control of materials for nanomedicine. Chem. Soc. Rev. 35 (11), $1095-1104$.

Favier, J., Revell, A., Pinelli, A., 2014. A Lattice Boltzmann-Immersed Boundary method to simulate the fluid interaction with moving and slender flexible objects. J. Comput. Phys. 261, 145-161.

Fedosov, D.A., Caswell, B., Karniadakis, G.E., 2010. A multiscale red blood cell model with accurate mechanics, rheology, and dynamics. Biophys. J. 98 (10), $2215-22225$

Fedosov, D.A., et al., 2010. Blood flow and cell-free layer in microvessels. Microcirculation 17 (8), 615-628.

Fedosov, D.A., et al., 2011. Multiscale modeling of red blood cell mechanics and blood flow in malaria. PLoS Comput. Biol. 7 (12), e1002270,

Fish, M.B., et al., 2017. Exploring deformable particles in vascular-targeted drug delivery: Softer is only sometimes better. Biomaterials 124 , $169-179$.

Godin, B., et al., 2012. Discoidal porous silicon particles: Fabrication and biodistribution in breast cancer bearing mice. Adv. Funct. Mater. 22 (20), $4225-4235$.

Guo, Z., Zheng, C., Shi, B., 2002. Discrete lattice effects on the forcing term in the lattice Boltzmann method. Phys. Rev. E (3) 65 (4 Pt $2 B$ ), 046308.

Kolhar, P., et al., 2013. Using shape effects to target antibody-coated nanoparticles to lung and brain endothelium. Proc. Natl. Acad. Sci. U.S.A. 110 (26), $10753-10758$ Kolonin, M.G., et al., 2004. Reversal of obesity by targeted ablation of adipose tissue. Nat. Med. 10 (6), 625-632.

Krüger, H., 2012. Computer Simulation Study of Collective Phenomena in Dense Suspensions of Red Blood Cells under Shear. Springer Science \& Business Media.

Lee, T.R., et al., 2013. On the near-wall accumulation of injectable particles in the microcirculation: smaller is not better. Sci. Rep. 3, 2079.

Li, X., et al., 2017. Biomechanics and biorheology of red blood cells in sickle cell anemia. J. Biomech. 50, 34-41.

Libby, P., 2002. Inflammation in atherosclerosis. Nature 420 (6917), 868-874.

Mody, N.A., King, M.R., 2007. Influence of Brownian motion on blood platelet flow behavior and adhesive dynamics near a planar wall. Langmuir 23 (11), 6321-6328.

Mulder, W.J., et al., 2014. Imaging and nanomedicine in inflammatory atherosclerosis. Sci. Transl. Med. 6 (239), 239 sr1.

Muller, K., Fedosov, D.A., Gompper, G., 2016. Understanding particle margination in blood flow - A step toward optimized drug delivery systems. Med. Eng. Phys. 38 (1), 2-10.

Mura, S., Nicolas, J., Couvreur, P., 2013. Stimuli-responsive nanocarriers for drug delivery. Nature Mater. 12 (11), 991-1003.

Muro, S., et al., 2008. Control of endothelial targeting and intracellular delivery of therapeutic enzymes by modulating the size and shape of ICAM-1-targeted carriers. Mol. Ther. 16 (8), 1450-1458.

Nakamura, M., Bessho, S., Wada, S., 2013. Spring-network-based model of a red blood cell for simulating mesoscopic blood flow. Int. J. Numer. Method Biomed. Eng. 29 (1), 114-128.

Neri, D., Bicknell, R., 2005. Tumour vascular targeting. Nat. Rev. Cancer 5 (6), 436-446.

Omori, T., et al., 2012. Tension of red blood cell membrane in simple shear flow. Phys. Rev. E 86 (5), 056321.

Palange, A.L., et al., 2017. Deformable discoidal polymeric nanoconstructs for the precise delivery of therapeutic and imaging agents. Mol. Ther. 25 (7), $1514-1521$,

Palomba, R., et al., 2018. Modulating phagocytic cell sequestration by tailoring nanoconstruct softness. ACS Nano.

Peer, D., et al., 2007. Nanocarriers as an emerging platform for cancer therapy. Nat. Nanotechnol. 2 (12), 751-760.

Pozrikidis, C., 2001. Effect of membrane bending stiffness on the deformation of capsules in simple shear flow. J. Fluid Mech. 440, $269-291$.

Pozrikidis, C., 2003. Numerical simulation of the flow-induced deformation of red blood cells. Ann. Biomed. Eng. 31 (10), $1194-1205$.

Pozrikidis, C., 2005. Axisymmetric motion of a file of red blood cells through capillaries. Phys. Fluids 17 (3), 031503.

Qian, Y.H., Dhumieres, D., Lallemand, P., 1992. Lattice Bgk models for Navier-Stokes equation. Europhys. Lett. 17 (6bis), $479-484$.

Roy, S., Qi, H.J., 2010. A computational biomimetic study of cell crawling. Biomech. Model. Mechanobiol. 9 (5), 573-581.

Shan, X.W., Yuan, X.F., Chen, H.D., 2006. Kinetic theory representation of hydrodynamics: a way beyond the Navier-Stokes equation. J. Fluid Mech. $413-441$.

Shi, L., Pan, T.-W., Glowinski, R., 2012. Deformation of a single red blood cell in bounded Poiseuille flows. Phys. Rev. E 85 (1), 016307.

Skalak, R., et al., 1973. Strain energy function of red blood cell membranes. Biophys. J. 13 (3), 245-264.

Soriano, A., et al., 2000. VCAM-1, but not ICAM-1 or MAdCAM-1, immunoblockade ameliorates DSS-induced colitis in mice. Lab. Invest. 80 (10), 1541-1551,

Sui, Y., et al., 2007. Transient deformation of elastic capsules in shear flow: effect of membrane bending stiffness. Phys. Rev. E 75 (6), 066301.

Sun, C., Migliorini, C., Munn, L.L., 2003. Red blood cells initiate leukocyte rolling in postcapillary expansions: a lattice Boltzmann analysis. Biophys. J. 85 (1), $208-225$.

Sun, C., Munn, L.L., 2008. Lattice Boltzmann simulation of blood flow in digitized vessel networks. Comput. Math. Appl. 55 (7), $1594-1600$.

Suzuki, K., Minami, K., Inamuro, T., 2015. Lift and thrust generation by a butterfly-like flapping wing-body model: immersed boundary-lattice Boltzmann simulations. J. Fluid Mech. 767, 659-695.

Ta, H.T., et al., 2018. The effects of particle size, shape, density and flow characteristics on particle margination to vascular walls in cardiovascular diseases. Expert Opin. Drug Deliv. 15 (1), 33-45.

Tan, J.F., et al., 2016. Characterization of nanoparticle dispersion in red blood cell suspension by the lattice Boltzmann-Immersed boundary method. Nanomaterials $6(2)$

Tan, J., Thomas, A., Liu, Y., 2011. Influence of red blood cells on nanoparticle targeted delivery in microcirculation. Soft Matter 8, 1934-1946.

de Tullio, M., Pascazio, G., 2016. A moving-least-squares immersed boundary method for simulating the fluid-structure interaction of elastic bodies with arbitrary thickness. J. Comput. Phys. 325, 201-225.

Vahidkhah, K., Bagchi, P., 2015. Microparticle shape effects on margination, near-wall dynamics and adhesion in a three-dimensional simulation of red blood cell suspension. Soft Matter 11 (11), 2097-2109.

Vanella, M., Balaras, E., 2009. A moving-least-squares reconstruction for embedded-boundary formulations. J. Comput. Phys. 228 (18), 6617-6628.

Wang, Y., et al., 2015. An immersed boundary-lattice Boltzmann flux solver and its applications to fluid-structure interaction problems. J. Fluids Struct. 54, $440-465$.

Ye, S.S., et al., 2014. Two-dimensional strain-hardening membrane model for large deformation behavior of multiple red blood cells in high shear conditions. Theor. Biol. Med. Model. $11(1), 19$.

Zhang, Q., Hisada, T., 2004. Studies of the strong coupling and weak coupling methods in FSI analysis. Internat. J. Numer. Methods Engrg. 60 (12), 2013-2029. Zou, Q.S., He, X.Y., 1997. On pressure and velocity boundary conditions for the lattice Boltzmann BGK model. Phys. Fluids 9 (6), $1591-1598$.

\end{document}